\documentclass[prl,nofootinbib,amsfonts]{revtex4}
\usepackage{graphicx,bm,amsmath,color}
\usepackage[latin1]{inputenc}

\newcommand{\be}{\begin{equation}}
\newcommand{\ee}{\end{equation}}
\newcommand{\xbf}{\bm{x}}
\newcommand{\kbf}{\bm{k}}

\begin{document}

\title{Quantum Noncanonical Field
Theory: Symmetries and Interaction}
\author{J.M. Carmona}
\email{jcarmona, cortes, indurain, dmazon@unizar.es}
\affiliation{Departamento de F\'{\i}sica Te\'orica,
Universidad de Zaragoza, Zaragoza 50009, Spain}
\author{J.L. Cort\'es}
\email{jcarmona, cortes, indurain, dmazon@unizar.es}
\affiliation{Departamento de F\'{\i}sica Te\'orica,
Universidad de Zaragoza, Zaragoza 50009, Spain}
\author{J. Indur\'ain}
\email{jcarmona, cortes, indurain, dmazon@unizar.es}
\affiliation{Departamento de F\'{\i}sica Te\'orica,
Universidad de Zaragoza, Zaragoza 50009, Spain}
\author{D. Maz\'on}
\email{jcarmona, cortes, indurain, dmazon@unizar.es}
\affiliation{Departamento de F\'{\i}sica Te\'orica,
Universidad de Zaragoza, Zaragoza 50009, Spain}
\begin{abstract}
The symmetry properties of a proposal to go beyond relativistic
quantum field theory based on a modification of the commutation
relations of fields are identified.
Poincar\'e invariance in an auxiliary spacetime is found in the
Lagrangian version of the path integral formulation.
This invariance is contrasted with the
idea of Doubly (or Deformed) Special Relativity (DSR). This analysis
is then used to go from the free theory of a complex field to an
interacting field theory.
\end{abstract}

\maketitle

\section{Introduction}

The validity of the framework of relativistic quantum field theory
(RQFT) at arbitrarily high energies has been put into question in the
last ten years by different approaches in the search for a quantum
theory of gravity~\cite{qugra}. In this context, the quantum theory of
noncanonical\footnote{What we refer in this paper as noncanonical
field has been named as noncommutative field in
Refs.~\cite{qncft,qncft2}. However, the extension of the concept to
fermion fields~\cite{neutrino} made more suitable the more rigorous
name of noncanonical field. This name fits also better with the
path integral formulation used in this paper.} fields was proposed
in Refs.~\cite{qncft,qncft2} as an
extension of RQFT. The free theory of noncanonical fields is
characterized by a modification of the canonical commutation relations
when quantizing in a deformed way a classical, relativistic Hamiltonian in the canonical
formalism. It is interesting that an explicit quantization of the
theory can be made in Fock space~\cite{qncft}, leading to a theory of
free particles with Lorentz non-invariant dispersion relations.

The idea of noncanonical fields has been implemented up to now at the
level of a free field theory.
The consistency of such a field theory when including interaction has
never been proved. In fact this is an ambitious program that should
start by investigating which is the proper criterion one should use to
introduce interaction terms in the theory. Since the theory breaks
Lorentz invariance, we no longer have special relativity as a symmetry
principle helping us to write interaction terms.

However, recently other symmetry principles aside from special relativity
have been considered. These symmetry principles preserve an energy
scale together with a velocity scale, and are grouped under the name
of Doubly (or Deformed) Special Relativity (DSR) \cite{dsr}. It might
be the case that a symmetry principle as or similar to DSR were
identified in the theory.

In this paper we explore this problem and try to define a way to go
beyond the free level for a theory of noncanonical fields. In
order to proceed, we will first identify
a symmetry principle replacing usual relativistic invariance in the
theory of a free noncanonical scalar field. Let us just remind the
main features of this theory (see Ref.~\cite{qncft} for more
details). It is defined by the Hamiltonian
\begin{eqnarray}
H &=& \int d^3 \xbf \,\, \mathcal{H}(\xbf), \nonumber \\
\mathcal{H}(\xbf) &=&
\Pi^{\dagger}({\xbf}) \Pi({\xbf}) \, + \,
\bm{\nabla}\Phi^{\dagger}({\xbf}) \bm{\nabla}\Phi({\xbf}) \, + \,
m^2 \Phi^{\dagger}({\xbf}) \Phi({\xbf}), \label{H}
\end{eqnarray}
together with the commutation relations
\begin{subequations}
\begin{eqnarray}
\left[\Phi(\xbf),\Phi^\dagger(\xbf')\right]&=&\theta \,\delta^3(\xbf-\xbf'), \\
\left[\Pi(\xbf),\Pi^\dagger(\xbf')\right]&=&0, \\
\left[\Phi(\xbf),\Pi^\dagger(\xbf')\right] &=& i
\,\delta^3(\xbf-\xbf')=\left[\Phi^\dagger(\xbf),\Pi(\xbf')\right], 
\end{eqnarray} \label{commrel}
\end{subequations}
where $\Phi$ is a complex scalar field and $\Pi$ is the momentum.
The previous commutation relations preserve rotational and
translational invariance in space, U(1) rigid symmetry in field space,
locality, and lead to RQFT at low energies ($\ll 1/\theta$).


The Heisenberg equations are obtained from $\partial_t
\Phi(\xbf)=-i[\Phi(\xbf),H]$ and $\partial_t
\Pi(\xbf)=-i[\Pi(\xbf),H]$~\cite{mand}. One then gets
\begin{equation}
\partial_t ^2 \Phi(\xbf)=(\bm{\nabla}^2-m^2)\Phi(\xbf) +
i[\theta(\bm{\nabla}^2-m^2)]\partial_t\Phi(\xbf).
\label{heisenberg}
\end{equation}
Writing $\Phi$ in momentum space
\begin{equation}
\Phi(\xbf,t)=\int \frac{d^4 k}{(2\pi)^4}\,  \Phi(k) e^{-ik\cdot x},
\end{equation}
a solution exists if and only if
\begin{equation}
\frac{k_0^2}{1+\theta k_0}-\kbf^2=m^2.
\label{disp}
\end{equation}
 It is then possible to find a representation of the field in a bosonic
Fock space:

\begin{equation}
\begin{split}
\Phi(\xbf,t)= &\int \frac{d^3 \kbf}{(2\pi)^3 }e^{i\kbf\cdot\xbf}
\left[\sqrt{\frac{E_a}{E_b(E_a+E_b)}} a_{\kbf}\,
e^{-i E_a t} \right.\\ + &
\left.\sqrt{\frac{E_b}{E_a(E_a+E_b)}}\, b_{-\kbf}^\dagger
e^{i E_b t}\right],
\end{split}
\end{equation}
where $a_{\kbf}, a_{\kbf}^\dagger, b_{\kbf}, b_{\kbf}^\dagger$ are two
kinds of bosonic annihilation and creation operators
$[ a_{\kbf}, a_{\kbf'}^\dagger] =[b_{\kbf}, b_{\kbf'}^\dagger] =
(2 \pi )^3 \delta ^3 \left( \kbf - \kbf'\right)$, and
$E_a=E_a(\kbf), E_b=E_b(\kbf)$ are the absolute values of the two solutions
for $k_0$ of the quadratic Eq.~(\ref{disp}),
\begin{eqnarray}
E_a(\kbf)&=&\omega_{\kbf}\left[\frac{1}{2}\theta\omega_{\kbf}+
 \sqrt{1+\frac{(\theta\omega_{\kbf})^2}{4}}\right]=k_0^+,
 \label{ena} \\
E_b(\kbf)&=&\omega_{\kbf}\left[-\frac{1}{2}\theta\omega_{\kbf}+
 \sqrt{1+\frac{(\theta\omega_{\kbf})^2}{4}}\right]=-k_0^- \, ,
 \label{enb}
\end{eqnarray}
with $\omega_{\kbf}=\sqrt{\kbf^2+m^2}$. The Hamiltonian and momentum operators can be
written (neglecting an infinite constant term) as
\begin{equation}
H=\int \frac{d^3 \kbf}{(2\pi)^3 } [E_a(\kbf)a_{\kbf}^\dagger a_{\kbf}
+ E_b(\kbf)b_{\kbf}^\dagger b_{\kbf}],
\label{HFock}
\end{equation}
\be
\bm{P} \,=\, \int \frac{d^3 \kbf}{(2\pi)^3 } \; \kbf \;
[a_{\kbf}^\dagger a_{\kbf} + b_{\kbf}^\dagger b_{\kbf}] \label{PFock},
\ee
showing that this is a theory of free particles of two types, with
energies $E_a(\kbf)$ and $E_b(\kbf)$, for each momentum $\kbf$.

In the $\theta\to 0$ limit, $E_a (\kbf) = E_b (\kbf) = \omega_{\kbf} $
corresponds to the particle-antiparticle degeneration of a
relativistic theory. The vacuum expectation value
of the time ordered product of field operators (propagator) of the
free theory is given by

\begin{equation}
\langle 0|T\left(\Phi (t,\xbf)\Phi ^\dagger (t',\xbf ')\right)|0\rangle=
\int \frac {d^4 k}{(2\pi)^4} ~e^{-ik_0(t-t')+i\kbf (\xbf - \xbf ')} ~
\dfrac{i\left( 1+\theta k_0\right)}
{\left( k_0 -E_a (\kbf)+i\epsilon \right)
 \left( k_0+E_b (\kbf)-i\epsilon \right) }
\label{propagator}\, .
\end{equation}

Note that we are breaking the discrete symmetry $\Phi \leftrightarrow
\Phi ^{\dagger}$ with the new symplectic
structure. The energies appearing in the Hamiltonian Eq.~(\ref{HFock})
are different functions of momentum and therefore $C$ transformation
is no longer a symmetry of the theory. $P$ and $T$ are still good
discrete symmetries, and therefore $CPT$ is broken.

According to Ref.~\cite{OWG2002}, Lorentz symmetry must be  broken in
the theory with $\theta\neq 0$. In fact, the dispersion
relation Eq.~(\ref{disp}) is not invariant under conventional Lorentz
transformations between inertial observers, though
it still satisfies rotational symmetry. In the following section
we will try to find out whether some kind of symmetry replacing Lorentz
invariance can still be defined for the noncanonical theory, which
would be helpful in the construction of an interacting theory.

\section{Symmetries of the theory of noncanonical fields}

The symmetries of a quantum field theory help us to understand its
spectrum. They give information about the relation between the energy
and momentum of the excitations over the ground state. In a free field
theory, transformations leaving the dispersion relation invariant are
transformations which connect solutions of the quantum equation of motion
among themselves. We will refer to them as symmetries of the
dispersion relation.

In RQFT the dispersion relation is Poincar\'e invariant.
However, the dispersion relation Eq.~(\ref{disp}) is not invariant under
conventional Lorentz transformations.
We start by considering whether a deformed transformation could be
defined so as to keep this  dispersion relation
invariant~\cite{TesisDiego}. A positive answer to this
question is the central issue in DSR theories. It is standard in DSR to
define auxiliary variables~\cite{visser,LSV05}
\begin{equation}
\kappa_0 = \frac{k_0}{\sqrt{1+\theta k_0}};\quad \bm{\kappa}=\kbf
\label{pi0}
\end{equation}
so that in terms of them the dispersion relation gets the form
\begin{equation}
\kappa_0^2-\bm{\kappa}^2 = m^2.
\end{equation}

Note that the previous relation between physical and auxiliary
momentum is only possible for $k_0 > -1/\theta$. But, looking at
Eq.~(\ref{disp}), we can see that, for all the solutions of the equation
motion, $k_0 > -1/\theta$. Then we will consider transformations on
functions of $k_\mu$ restricted to this domain.

The auxiliary variables satisfy the usual, Lorentz invariant,
dispersion relation. We can then define the generators of boosts as
the usual expressions
\begin{equation}
N_i = i(\kappa_0 \partial_{\kappa_i}+\kappa_i \partial_{\kappa_0}),
\label{genaux}
\end{equation}
so that in terms of the original variables
\be
N_i=i\left[\frac{k_0}{\sqrt{1+\theta k_0}}\partial_{k_i}+
\frac{(1+\theta k_0)^{3/2}}{1+\theta k_0/2} k_i\partial_{k_0}\right], \label{boost}
\ee
which leaves Eq.~(\ref{disp}) invariant by construction.

The algebra of the generators of boosts is not modified with respect
to the usual Lorentz algebra, as one can easily check by writing the
generators in terms of the auxiliary variables,
Eq.~(\ref{genaux}). This is not the case, however, of the algebra of
boost generators with the four-momentum, which is different
from the usual Poincar\'e algebra, and reduces to it in the $\theta\to
0$ limit,
\be
[N_i,k_0]=i k_i \frac{(1+\theta k_0)^{3/2}}{1+\theta k_0/2}\, ,\quad
[N_i,k_j]=i \delta_{ij} \frac{k_0}{\sqrt{1+\theta k_0}}.
\label{nl}
\ee
Poincar\'e algebra has been, therefore, deformed to a new closed,
non-linear, algebra.

Our theory, however, is a quantum theory of fields. This means that,
once we have identified a principle of relativity for the invariance
of the dispersion relation of particles, it is necessary to find the
corresponding representation of the symmetry in the Fock space.

Every vector in the Fock space can be written as a sum of tensor products
of vectors belonging to the single-particle Hilbert space. In our case
we have two kinds of single-particle spaces, that is, those corresponding
to particles of types $a$ and $b$, respectively. We will define a
representation of Lorentz transformations in each of these spaces by
saying how they act on a basis of vectors. The basis elements are the
states with a given momentum $\kbf$
\be
|\kbf\rangle_a = \sqrt{2\omega_{\kbf}} a^\dagger_{\kbf} |0\rangle; \quad
|\kbf\rangle_b = \sqrt{2\omega_{\kbf}} b^\dagger_{\kbf} |0\rangle,
\ee
for the corresponding spaces of particles $a$ and $b$.
We are using the same normalization factors of the relativistic
theory; in fact the spatial components of the auxiliary
four-momentum variables Eq.~(\ref{pi0}) transforming linearly under Lorentz
transformations  are just the components of the momentum
$\kbf$.

A Lorentz transformation $\Lambda$ is represented in the Fock space by
a unitary operator $U_{\Lambda}$, according to
\be
|\kbf'\rangle_a = U_{\Lambda}|\kbf\rangle_a \,; \quad
|\kbf'\rangle_b = U_{\Lambda}|\kbf\rangle_b
\ee
where $\kbf'$ is the momentum of the state
which results of applying a standard Lorentz transformation on a state with
momentum $\kbf$ and energy $\omega_{\kbf}$ in special relativity. Then Lorentz
transformations in the Fock space are just those of RQFT
\begin{eqnarray}
U_{\Lambda} a^\dagger_{\kbf} U_{\Lambda}^\dagger &=&
\sqrt{\frac{\omega_{\kbf'}}{\omega_{\kbf}}} a^\dagger_{\kbf'}, \label{Ua}\\
U_{\Lambda} b^\dagger_{\kbf} U_{\Lambda}^\dagger &=&
\sqrt{\frac{\omega_{\kbf'}}{\omega_{\kbf}}} b^\dagger_{\kbf'}. \label{Ub}
\end{eqnarray}

Any state in the Fock space can be written as a linear combination of
states which result from acting with a product of $a^\dagger$,
$b^\dagger$ operators on the vacuum. Then Eqs. (\ref{Ua}-\ref{Ub}) define
the transformation of any state in the Fock space. Once we have
identified a unitary representation of the Lorentz transformations in
the Fock space we can work out the transformations of the Hamiltonian
Eq.~(\ref{HFock}) and the momentum operator Eq.~(\ref{PFock}).

Under Lorentz transformations,
\begin{eqnarray}
U_{\Lambda} H U_{\Lambda}^\dagger \,&=&\, \int \frac{d^3 \kbf'}{(2\pi)^3
  } \left[ E_a(\kbf) a_{\kbf'}^\dagger a_{\kbf'}
+
E_b(\kbf) b_{\kbf'}^\dagger b_{\kbf'}\right], \\
U_{\Lambda} \bm{P} U_{\Lambda}^\dagger \,&=&\, \int \frac{d^3 \kbf'}{(2\pi)^3
  } \; \kbf \; \left[ a_{\kbf'}^\dagger a_{\kbf'} +
b_{\kbf'}^\dagger b_{\kbf'} \right].
\end{eqnarray}

There is no way to write the transformed operators as a function of
the Hamiltonian and the momentum operator so that there is no closed
deformed Poincar\'e algebra including the Hamiltonian and the momentum
as generators. This can be seen from the following argument. The effect of the
transformation of the
Hamiltonian on the Fock space representation is to replace the
coefficients $E_a(\kbf')$, $E_b(\kbf')$
of $a_{\kbf'}^\dagger a_{\kbf'}$, $b_{\kbf'}^\dagger b_{\kbf'}$ by
$E_a(\kbf)$, $E_b(\kbf)$. Similarly the effect on the momentum
operator is to replace the coefficient $\kbf'$ of $a_{\kbf'}^\dagger
a_{\kbf'}$, $b_{\kbf'}^\dagger b_{\kbf'}$ by $\kbf$. Although there is
a nonlinear relation between the coefficients in the original
and transformed operators, the linear dependence on $a^\dagger a$,
$b^\dagger b$ prevents us to identify a nonlinear closed algebra.

However, if we project on the one particle sector the
transformation laws of the Hamiltonian and the momentum operator
we recover the nonlinear deformation of the Poincar\'e algebra
Eq.~(\ref{nl}). The energy-momentum relations for the particles of the free
noncanonical theory can then be understood as a consequence of nonlinear
deformations of the Poincar\'e algebra depending on the noncommutativity
length scale $\theta$.

The role of the ultraviolet (UV) parameter $\theta^{-1}$, as well as
the deformation of relativistic invariance, can be compared with the
typical situation in DSR theories.
Eq.~(\ref{enb}) shows that when $|\kbf|\to \infty$ , then
$E_b\to \theta^{-1}$. This and the fact that $E_b (\kbf)$ is
a monotonically increasing function means that $\theta^{-1}$ is
the supremum of the energy of particle $b$. Since Eq.~(\ref{disp})
is invariant under the deformed boosts generated by $N_i$ in
Eq.~(\ref{boost}), every inertial observer will agree on this supreme
value for $E_b$.

This fact gives a direct physical meaning to the
parameter $\theta$ in the theory of noncanonical fields, in analogy
to the similar role played by the Planck mass, $M_P$, in DSR theories.
However, there is not an observer-independent energy scale associated
with particle $a$. Therefore, we have a DSR 3 (i.e,
an observer-independent energy cutoff)  realization in the
one-`particle $b$' sector and a smoothly modified special relativity
(i.e. without any cutoff in energy or momentum) in the one-`particle $a$'
sector~\cite{lukierski}. The same conclusions can
be obtained from the expression for the auxiliary variables. One has a
one to one mapping Eq.~(\ref{pi0})
between $-\infty < \kappa_0 < \infty$ and $-1/\theta < k_0 <
\infty$.

There is no extension of the deformed Poincar\'e symmetry of the
single-particle sector to the multiparticle sectors of the Fock
space. In
this sense we can say that although we can find  a parallelism between
the introduction of an UV scale through noncommutativity of field
operators and the attempts to make the relativity principle
compatible with a new invariant energy scale (DSR) in the one
particle sector, the quantum theory of noncanonical fields is
not a quantum field theory realization of DSR.

Another way to arrive to the same conclusion is by contrasting the
trivial additive composition law of energy and momentum in the free
theory of noncanonical fields with the necessity for considering a
nontrivial composition law in DSR, owing to the nonlinear
representation in momentum space of Lorentz transformations.

\section{Path integral formulation of the free theory}

In order to apply our understanding from the point of view of
symmetries of the free noncanonical field theory to the introduction
of interactions it is convenient to use the path integral
formulation of field theory. In this section we translate the standard
derivation~\cite{Weinberg} of the path integral approach of a quantum
theory to the free noncanonical field theory, which reduces to finding a
representation of the scalar propagator
\be
\langle 0|T\left(\Phi(x_A)
\Phi^\dagger(x_B)\right)|0\rangle
\ee
as a path integral.

First we introduce the basis of eigenstates
$|\pi(\xbf);t\rangle$
\be
\Pi(t,\xbf)|\pi(\xbf);t\rangle = \pi(\xbf) |\pi(\xbf);t\rangle
\ee
where $\Pi(t,\xbf)$ are the Heisenberg-picture operators
\be
\Pi(t,\xbf) = e^{iHt} \Pi(\xbf) e^{-iHt}.
\ee
This allows to express the scalar propagator as
\be
\int \left[\prod_{\xbf} d\pi'(\xbf)\right] \left[\prod_{\xbf}
  d\pi(\xbf)\right] \; \langle 0|\pi'(\xbf);t_f\rangle \;
\langle\pi'(\xbf);t_f|T\left(\Phi(x_A)
\Phi^{\dagger}(x_B)\right)|\pi(\xbf);t_i\rangle \;
\langle\pi(\xbf);t_i|0\rangle.
\ee

Owing to the noncommutativity of fields there are no eigenstates of the
complex scalar field operator. But it is possible to introduce the
linear combinations
\be
\Phi_c(t,\xbf) = \Phi(t,\xbf) - \frac{i\theta}{2} \Pi(t,\xbf)
\ee
and the basis of eigenstates $|\phi_c(\xbf);t\rangle$
\be
\Phi_c(t,\xbf)|\phi_c(\xbf);t\rangle = \phi_c(\xbf) |\phi_c(\xbf);t\rangle.
\ee
Note that $\Phi_c$ and $\Pi$ form a canonical conjugate pair of
variables.

Then one can repeat step by step the standard derivation of the path
integral formulation~\cite{Weinberg}.
First one can use the completeness conditions
\begin{eqnarray}
\int \left[\prod_{\xbf} d\pi(\xbf)\right] \;  |\pi(\xbf);t\rangle
\langle \pi(\xbf);t| \; &=& 1 \\
\int \left[\prod_{\xbf} d\phi_c(\xbf)\right] \; |\phi_c(\xbf);t\rangle
\langle \phi_c(\xbf);t| \; &=&  1
\end{eqnarray}
for all $t$ with $t_i<t<t_f$ to write the matrix elements of the time
ordered product of two field operators between eigenstates of
$\Pi(t_i, \xbf)$ and $\Pi(t_f, \xbf)$ as a path integral\footnote{The
arbitrariness in the normalization of the integration measure leaves an
undetermined proportionality factor in all equations involving a path
integral.}
\begin{multline}
\langle\pi'(\xbf);t_f|T\left(\Phi(x_A)
\Phi^{\dagger}(x_B)\right)|\pi(\xbf);t_i\rangle \propto
\displaystyle\int_{\begin{subarray}{l}\pi(t_i,\xbf)=\pi(\xbf)
    \\\pi(t_f,\xbf)=\pi'(\xbf)\end{subarray}}
\left[\prod_{t,\xbf} d\pi(t,\xbf)\right] \left[\prod_{t,\xbf}
  d\phi(t,\xbf)\right] \phi(x_A) \phi^{*}(x_B) \\
 \;\; \exp\left[-i\int_{t_i}^{t_f} dt\left\lbrace\int d^3x
  \left(\phi(t,\xbf) \partial_t\pi^{*}(t,\xbf) - \frac{i\theta}{2}\pi(t,\xbf)
\partial_t\pi^{*}(t,\xbf) + \text{c.c.}\right) + H[\pi(t),\phi(t)]
\right\rbrace \right]\, ,
\end{multline}
where we have used the new integration variable $\phi (x)\equiv \phi
_c (x) + i\frac{\theta}{2} \pi (x)$ instead of $\phi_c (x)$~\cite{CA01}.

Next one can calculate, from the Fock space representation of the free
noncanonical theory discussed in the Introduction section, the projection of
the vacuum on the eigenstates $|\pi(\xbf)\rangle$
\be
\langle \pi(\xbf) | 0\rangle = N \exp\left(-\frac{1}{2} \int d^3y d^3z
  \Delta(\bm{y}, \bm{z}) \pi^{*}(\bm{y}) \pi(\bm{z})\right),
\ee
where $N$ is a normalization constant and
\be
\Delta (\bm{y}, \bm{z}) \,=\, \int \frac{d^3k}{(2\pi)^3} \;
\left(\frac{E_a(\kbf)+E_b(\kbf)}{\omega_{\kbf}^2}\right) \, e^{i\kbf(\bm{y}-\bm{z})}.
\ee
Using the relation
\be
f(\infty) + f(-\infty) = \lim_{\epsilon\to 0^+} \epsilon
\int_{-\infty}^{\infty} dt \, f(t) e^{-\epsilon|t|}
\ee
one has
\be
\lim_{\substack{t_i\to -\infty\\t_f\to \infty}}
\langle 0|\pi'(\xbf);t_f\rangle \;
\langle\pi(\xbf);t_i|0\rangle \; = |N|^2
\exp\left(-\frac{1}{2}\epsilon \int d^3y d^3z \int_{-\infty}^{\infty}
dt \Delta(\bm{y}, \bm{z}) \pi^{*}(t,\bm{y}) \pi(t,\bm{z})
e^{-\epsilon|t|} \right) .
\ee

The final result for the propagator is
\begin{multline}
\langle 0|T\left(\Phi(x_A)
  \Phi^\dagger(x_B)\right)|0\rangle \propto
\int \left[\prod_{t,\xbf} d\pi(t,\xbf)\right] \left[\prod_{t,\xbf}
  d\phi(t,\xbf)\right] \phi(x_A) \phi^{*}(x_B) \\
\exp\left[-i\int_{-\infty}^{\infty} dt\left\lbrace\int d^3x
  \left(\phi(t,\xbf) \partial_t\pi^{*}(t,\xbf) - \frac{i\theta}{2}\pi(t,\xbf)
\partial_t\pi^{*}(t,\xbf) + \text{c.c.}\right) + H[\pi(t),\phi(t)] \right.\right.\\
\left.\left. -
  \frac{1}{2} i\epsilon \int d^3y d^3z \Delta(\bm{y}, \bm{z})
  \pi^{*}(t,\bm{y}) \pi(t,\bm{z}) e^{-\epsilon|t|} \right\rbrace \right].
\end{multline}

In order to get the Lagrangian\footnote{Strictly speaking we
should refer to this as a generalized Lagrangian formulation
since we are not using canonical field variables.} version of the path
integral formulation one has to integrate over $\pi$. Since the
argument of the exponential is quadratic in $\pi$ the integral will be
proportional to the integrand evaluated at the stationary point of its
argument. Then one has
\be
\langle 0|T\left(\Phi(x_A)
\Phi^\dagger(x_B)\right)|0\rangle \propto
\int \left[\prod_{t,\xbf} d\phi(t,\xbf)\right] \phi(x_A) \phi^{*}(x_B)
e^{i S[\phi]}
\ee
with\footnote{We have assumed field configurations such that total
  derivatives in the generalized Lagrangian density can be ignored.}
\be
S[\phi] \,=\, \int dt \, d^3x \, \phi^{*}(t,\xbf)
\left[-\frac{\partial_t^2}{1+i\theta \partial_t-i\epsilon}+
\bm{\nabla}^2 -m^2\right] \phi(t,\xbf).
\label{S}
\ee
Because we are expressing the action as a functional of noncanonical
variables, inverses of differential operators appear in the
action. A more explicit way of writing the action is

\be
S[\phi] \,=\, \int dt \, dt' \, d^3x \, \phi^{*}(t,\xbf)
\left[-D(t-t')\partial_{t'}^2+
\delta(t-t')(\bm{\nabla}^2 -m^2)\right] \phi(t',\xbf),
\ee
where

\be
D(t-t')=\int \frac{d\omega}{2\pi}
\frac{e^{-i\omega(t-t')}}{(1+\theta\omega -i\epsilon)}; 
\ee
in this expression, the invariance under time translations is manifest. The
 invariance under time translations is in agreement with the conservation
 of energy that can be deduced from the explicit time independence of
 the Hamiltonian \eqref{H}.
 At this point we can introduce the generating
functional of Green functions
\be
Z[j] \propto \int \left[\prod_{t,\xbf} d\phi(t,\xbf)\right]
e^{i S[\phi] + i \int d^4x \,
  \left(j^{*}(x)\phi(x) + j(x)\phi^{*}(x)\right)},
\label{Z}
\ee
where once more the argument in the exponential is quadratic in
$\phi$. Then $Z[j]$ is an exponential with an argument
quadratic in $j$. From this result one can calculate the two-point
Green function from the generating functional $Z[j]$
\begin{eqnarray}
G^{(2)}(t_A, \xbf_A; t_B, \xbf_B) &=& - \frac{1}{Z[j]}\left.\frac{\delta^2 Z[j]}{\delta
j^{*}(t_A, \xbf_A) \delta j(t_B, \xbf_B)}\right|_{j=0} \nonumber \\
&=&\int \frac {d^4 k}{(2\pi)^4}
~e^{-ik_0(t_A-t_B)+i\kbf (\xbf_A - \xbf_B)} ~ \dfrac{i\left( 1+\theta k_0\right)}
{\left( k_0 -E_a (\kbf)+i\epsilon \right)
 \left( k_0+E_b (\kbf)-i\epsilon \right) }
\label{G}
\end{eqnarray}
and one can check that it coincides with the result for the
vacuum expectation value of the time ordered product of field
operators in the operator formalism Eq.~(\ref{propagator}).

The generating functional $Z[j]$ in Eq.~(\ref{Z}) provides a formalism for
the free noncanonical theory as an integral in a complex field
configuration space.

\subsection{Map to the relativistic free field theory}

The effect of the noncommutativity of fields in the path integral
formulation can be seen through the $\theta$-dependence of the action
Eq.~(\ref{S}). If we introduce
\be
\phi(k_0, \xbf) \,=\, \int dt \, e^{ik_0 t} \, \phi(t, \xbf)
\ee
then the action can be written as
\be
S[\phi] \,=\, \int \frac{dk_0}{2\pi} \int d^{3}\xbf \, \phi^{*}(k_0, \xbf)
\left[\frac {k_0^2}{1+\theta
k_0-i\epsilon} + \bm{\nabla}^2 - m^2\right] \phi(k_0, \xbf).
\ee
At this point it is convenient to split the integration on $k_0$ into
two pieces
\be
S[\phi] \,=\, S^{\theta}[\phi^{\theta}] + {\bar S}^{\theta}[{\bar
    \phi}^{\theta}]
\label{S-Sbar}
\ee
with
\begin{eqnarray}
\phi^{\theta}(t, \xbf) &=& \int_{-1/\theta}^{\infty} dk_0
 \, e^{- ik_0 t} \, \phi(k_0, \xbf), \\
{\bar \phi}^{\theta}(t, \xbf) &=& \int_{-\infty}^{-1/\theta} \frac{dk_0}{2\pi}
 \, e^{- ik_0 t} \, \phi(k_0, \xbf).
\end{eqnarray}
In the first piece ($k_0>-1/\theta$) one can make a change of variables
\be
\kappa_0 = \frac{k_0}{\sqrt{1+\theta k_0}}
\ee
and introduce a new field variable
\begin{eqnarray}
\phi_r(\kappa_0, \xbf) &=& \sqrt{\frac{dk_0}{d\kappa_0}} \, \phi(k_0,
\xbf) \\
\phi_r(\tau, \xbf) &=& \int_{-\infty}^{\infty} \frac{d\kappa_0}{2\pi} \,
e^{-i\kappa_0\tau} \, \phi_r(\kappa_0, \xbf),
\label{phir}
\end{eqnarray}
which leads to
\be
S^{\theta}[\phi^{\theta}] \,=\, S_r[\phi_r],
\label{Stheta-Sr}
\ee
where $S_r$ is the action of the relativistic theory of the free
complex scalar field $\phi_r$
\be
S_r[\phi_r] = \int_{-\infty}^{\infty} d\tau \, d^{3}\xbf \,
\phi_r^{*}(\tau, \xbf) \left[-\partial_{\tau}^2 + \bm{\nabla}^2 - m^2 +
  i\epsilon\right] \phi_r (\tau, \xbf).
\ee

The relation Eq.~(\ref{Stheta-Sr}) is the translation to the path integral
formulation of the possibility to find a nonlinear change of
variables which brings the dispersion relation of the
noncanonical theory into the relativistic dispersion relation.
This nonlinear change of variables was the starting point to identify
a deformed Poincar\'e invariance in the one particle sectors including
the Hamiltonian as one of the generators.

Alternatively, it is possible to identify a Poincar\'e
group of symmetries of the action (i.e.: a group of
transformations which leave the action invariant).
To see this one has to consider a
transformation on the complex field  leaving invariant the component
$\bar{\phi}^{\theta}$ and translating the Poincar\'e transformation of
the relativistic complex scalar field $\phi_r$ to the related component
$\phi^{\theta}$. The action $S[\phi]$ is obviously invariant under
such transformations because it is a sum of a contribution
$S^{\theta}[\phi^{\theta}]$, which is invariant due to the relation
Eq.~(\ref{Stheta-Sr}) and the Poincar\'e invariance of the relativistic
action, and a contribution ${\bar S}^{\theta}[{\bar \phi}^{\theta}]$
depending only on the field component which is invariant under the
transformation. One of the generators of these transformations
corresponds to the translation in the time argument $\tau$ of the
variable $\phi_r$,
which has nothing to do with a translation in the time argument $t$ of
the field $\phi$ generated by the Hamiltonian of the free
noncanonical field theory.
The effect of the group of symmetries of the action on
the variable $\phi_r$ is
\be
\phi'_r(\tau'_{(\Lambda,a)}, \xbf'_{(\Lambda,a)}) = \phi_r(\tau,\xbf) \, ,
\ee
where $(\tau'_{(\Lambda,a)}, \xbf'_{(\Lambda,a)})$ is the ordinary
Poincar\'e transformation of the Minkowskian four vector
$(\tau,\xbf)$. Therefore the spacetime realization of the identified
Poincar\'e group of symmetries of the action requires the use of the auxiliary
spacetime formed by the elements $\{(\tau,\xbf)\}$.
We are identifying a Poincar\'e invariance of the
path integral formulation by considering a non-conventional
representation of Poincar\'e transformations on the space of
configurations of the noncanonical field variables. It may appear
bizarre to identify a Poincar\'e symmetry which is not a symmetry of
the physical spacetime. In order to interpret this symmetry as a
spacetime symmetry an auxiliary time variable $\tau$ must be
introduced. This group of symmetries also leaves the
 equal-time commutation relations \eqref{commrel} invariant. If these
transformations could be interpreted as changes of reference frame, then 
the commutation relations would not select a preferred reference frame.
 This mathematical trick will prove to be useful in order
to study the renormalization of the theory.

At the level of the generating functional of Green functions $Z[j]$
one can introduce a similar decomposition of the sources
\be
j(t, \xbf) = j^{\theta}(t,\xbf) + {\bar j}^{\theta}(t,\xbf)
\ee
with
\begin{eqnarray}
{\bar j}^{\theta}(t,\xbf) &=& \int_{-\infty}^{-1/\theta}
\frac{dk_0}{2\pi} \, e^{-ik_0 t}
\, j(k_0, \xbf), \\
j^{\theta}(t,\xbf) &=& \int_{-1/\theta}^{\infty} \frac{dk_0}{2\pi} \, e^{-ik_0t}
\, j(k_0, \xbf).
\end{eqnarray}
One gets
\be
Z[j] \,=\, Z_r[j_r] \; {\bar Z}^{\theta}[{\bar j}^{\theta}],
\label{Zsplit}
\ee
where $Z_r[j_r]$ is the generating functional of Green functions of the
relativistic free theory of a complex field with
\be
j_r(\tau, \xbf) = \int dt \, K_{\theta}(\tau, t) \, j^{\theta}(t, \xbf) \label{tauxt}
\ee
and
\be
K_{\theta}(\tau, t) \,=\,  \int_{-\infty}^{\infty}
\frac{d\kappa_0}{2\pi} \; e^{i(t k_0 - \tau\kappa_0)} \;
\sqrt{\frac{dk_0}{d\kappa_0}}\,, \label{kern}
\ee
with
\be
k_0 = \kappa_0 \left[\sqrt{1+\left(\frac{\theta\kappa_0}{2}\right)^2} +
  \frac{\theta\kappa_0}{2}\right] \, .
\ee

Eq.~(\ref{kern}) gives the explicit relation between the physical time
$t$ and the auxiliary time $\tau$ through the correspondence
Eq.~(\ref{tauxt}) between functions defined in the physical spacetime and
functions defined in the auxiliary spacetime.

Then one has
\begin{eqnarray}
(-i)\frac{\delta Z[j]}{\delta j(t, \xbf)} &=& {\bar Z}^{\theta}[{\bar
    j}^{\theta}] \; \int dt'\int_{-1/\theta}^{\infty} \frac{dk_0}{2\pi}
e^{-ik_0(t-t')} \int d\tau K_{\theta}(\tau, t')
(-i)\frac{\delta Z_r[j_r]}{\delta j_r(\tau, \xbf)} \nonumber \\
&+& Z_r[j_r] \; \int dt'\int_{-\infty}^{-1/\theta} \frac{dk_0}{2\pi}
e^{-ik_0(t-t')} (-i)\frac{\delta {\bar Z}^{\theta}[{\bar
    j}^{\theta}]}{\delta {\bar j}^{\theta}(t', \xbf)}.
\end{eqnarray}
The splitting of the generating functional in Eq.~(\ref{Zsplit}) just
corresponds at the level of the propagator to the separation of the
contributions of modes with $k_0>-1/\theta$ and those with $k_0<-1/\theta$,

\begin{multline}
G^{(2)}(t_A, \xbf_A; t_B, \xbf_B)=
\int dt'_A dt'_B \int_{-1/\theta}^{\infty} \frac{dk_{A0}}{2\pi}
\frac{dk_{B0}}{2\pi}
e^{ik_{A0}(t_A-t'_A)} e^{-ik_{B0}(t_B-t'_B)}
\int d\tau_A d\tau_B K_{\theta}^{*}(\tau_A, t'_A)
K_{\theta}(\tau_B, t'_B)
 \\ G^{(2)}_r(\tau_A, \xbf_A; \tau_B, \xbf_B) + \int
dt'_A dt'_B \int_{-\infty}^{-1/\theta}
\frac{dk_{A0}}{2\pi} \frac{dk_{B0}}{2\pi} e^{ik_{A0}(t_A-t'_A)}
e^{-ik_{B0}(t_B-t'_B)}
{\bar G}^{\theta}(t'_A, \xbf_A; t'_B, \xbf_B)
\label{G2}\, ,
\end{multline}
where
\begin{eqnarray}
{\bar G}^{\theta}(t_A,\xbf_A; t_B, \xbf_B) &=&
- \frac{1}{{\bar Z}^{\theta}[j]}\left.\frac{\delta^2 {\bar
Z}^{\theta}[j]}{\delta {\bar j}^{\theta *}(t_A, \xbf_A) \delta {\bar
j}^{\theta}(t_B, \xbf_B)}\right|_{{\bar j}^{\theta}=0} \nonumber
\\&=&\int^{-1/\theta}_{-\infty} \frac {d k_0}{2\pi}\int \frac {d^3
  \kbf}{(2\pi)^3} ~e^{-ik_0(t_A-t_B)+i\kbf (\xbf_A - \xbf_B)} ~
\dfrac{i\left( 1+\theta k_0\right)}
{\left( k_0 -E_a (\kbf) \right)
 \left( k_0+E_b (\kbf) \right) } \, .
\label{Gtheta}
\end{eqnarray}
The expression for the propagator Eq.~(\ref{G2}) is just the result of
considering a second derivative of the generating functional $Z[j]$
with respect to the sources $j(t,\xbf)$ together with the
decomposition of the source as a sum of two independent components
$j^{\theta}$, ${\bar j}^{\theta}$. Note that Eq.~(\ref{Gtheta}) is just
the free propagator Eq.~(\ref{G}) restricted to modes with
$k_0 < -1/\theta$ where the $i\epsilon$ factors can be omitted owing to
the factor $1+\theta k_0$ in the numerator.

\section{Interaction}

In order to go beyond the free theory of a complex
noncanonical field including interactions, a first attempt could be based
on the addition of  a term proportional to $[ \Phi^{\dagger}({\xbf}) \Phi({\xbf})]^2$
in the Hamiltonian density Eq.~(\ref{H}). If one goes to the path integral
formulation following the same steps of the free theory one finds that
the relation with the relativistic theory found at the level of the
free theory is lost.
A consistent generalization
of the standard perturbative analysis of the relativistic theory is
problematic. The Poincar\'e invariance of the free theory is lost and,
lacking a characterization in terms of symmetries of the interacting
theory, renormalizability will also be lost.

In a second attempt to go beyond the free theory one could take
the Lagrangian version of the path integral formulation of the
free noncanonical free theory Eq.~(\ref{S}) as a starting
point, and then add directly at this
level a term proportional to $(\phi^{*}\phi)^2$. In this case too one
looses the relation with the relativistic theory found at the level of the
free theory.
This is due to the fact that a decomposition of the
field into modes with $k_{0}>-1/\theta$ and $k_{0}< -1/\theta$ does
not lead in this case to a splitting of the action (and the generating
functional) because the non-quadratic terms couple the two types of
modes. Then the Poincar\'e invariance of the free theory is lost.

In order to consider a theory with interactions maintaining the
Poincar\'e invariance of the free theory it is necessary to keep the splitting of the
action Eq.~(\ref{S-Sbar}) also in the interacting theory. The
Poincar\'e invariance restricts the possible nonquadratic terms in the
first contribution to the action through Eq.~(\ref{Stheta-Sr}), but not in
the second one. A simple way to achieve a perturbatively
renormalizable interacting theory is to
introduce the interaction only through a term proportional to
$(\phi_r^{*}\phi_r)^2$ in $S_r$.
One has then a splitting at the
level of the generating functional of Green functions, a separation of
modes with $k_0>-1/\theta$ and $k_0<-1/\theta$, and a simple relation
of the Green functions of the noncanonical field theory and those of
the relativistic theory. Then the multiplicative renormalizability of
the relativistic field theory can be translated to the noncanonical
theory.

In fact the relation between the propagator of the noncanonical
field theory and the relativistic propagator Eq.~(\ref{G2}) of the free
theory applies also in the interacting theory.
All the contributions from modes with $k_0 <
-1/\theta$ can be expressed in terms of the function
$\bar{G}^\theta$. The contribution from modes with $k_0 >
-1/\theta$ to a given $n$-point Green function of the quantum
noncanonical field theory, $G^{(n)}$, can be expressed in terms of
the $m$-point Green functions $G_r^{(m)}$ of the relativistic theory
with interaction (with $m\leq n$).
In the case of the four-point Green function one has
\begin{eqnarray}
&& {\hskip 4cm} G^{(4)}(t_A, \xbf_A; t_B, \xbf_B; t_C, \xbf_C; t_D,
  \xbf_D)= \nonumber \\
&&\int dt'_A dt'_B dt'_C dt'_D
\int_{-1/\theta}^{\infty} dk_{A0} dk_{B0} dk_{C0} dk_{D0}
\; e^{ik_{A0}(t_A-t'_A)} e^{ik_{B0}(t_B-t'_B)}
e^{-ik_{C0}(t_C-t'_C)} e^{-ik_{D0}(t_D-t'_D)} \nonumber \\
&&\int d\tau_A d\tau_B d\tau_C d\tau_D \; K_{\theta}^{*}(\tau_A, t'_A)
K_{\theta}^{*}(\tau_B, t'_B) K_{\theta}(\tau_C, t'_C)
K_{\theta}(\tau_D, t'_D) \; G^{(4)}_r(\tau_A, \xbf_A; \tau_B, \xbf_B;
\tau_C, \xbf_C; \tau_D, \xbf_D) \nonumber \\
&&+ \int dt'_A dt'_B dt'_C dt'_D
\left[\int_{-1/\theta}^{\infty} dk_{A0} dk_{C0}
\int_{-\infty}^{-1/\theta} dk_{B0} dk_{D0}
\; e^{ik_{A0}(t_A-t'_A)} e^{ik_{B0}(t_B-t'_B)}
e^{-ik_{C0}(t_C-t'_C)} e^{-ik_{D0}(t_D-t'_D)}\right. \nonumber \\
&&\left.\int d\tau_A d\tau_C \; K_{\theta}^{*}(\tau_A, t'_A)
K_{\theta}(\tau_C, t'_C) \; G^{(2)}_r(\tau_A, \xbf_A; \tau_C, \xbf_C)
{\bar G}^{\theta}(t'_B, \xbf_B; t'_D, \xbf_D) + \text{3
  permutations}\right] \nonumber \\
&&+ \int dt'_A dt'_B dt'_C dt'_D
\int_{-\infty}^{-1/\theta} dk_{A0} dk_{B0} dk_{C0} dk_{D0}
\; e^{ik_{A0}(t_A-t'_A)} e^{ik_{B0}(t_B-t'_B)}
e^{-ik_{C0}(t_C-t'_C)} e^{-ik_{D0}(t_D-t'_D)} \nonumber\\
&& {\hskip 2cm} \left[{\bar G}^{\theta}(t'_A, \xbf_B; t'_C, \xbf_D)
{\bar G}^{\theta}(t'_B, \xbf_B; t'_D, \xbf_D) +
\text{permutation}\right].
\label{G4}
\end{eqnarray}
Noticeably for the fully connected Green functions one gets a simpler expression with just the first term on the right hand side of
Eq.~(\ref{G4}), and modes with $k_0<-1/\theta$ are irrelevant.

All these relations are a consequence of the correspondence
between actions Eq.~(\ref{Stheta-Sr}) and the fact that ${\bar
Z}^{\theta}$ is an exponential with an argument which is quadratic
in the sources (we do not consider non-quadratic terms in the
field components with $k_{0}<-1/\theta$). The inclusion of
interaction terms compatible with renormalizability in the
$\bar{S}^\theta$ contribution to the action would require a
generalization of the standard arguments (power counting,
symmetries) to this contribution.

Since we have introduced the interaction directly at the level of the
Lagrangian version of the path integral formulation there is no clear
way to identify the Hamiltonian of the interacting theory and there is
no clear physical interpretation of the theory defined by the Green
functions $G^{(n)}(x_1, x_2,..., x_n)$. In particular it is not
possible to relate these objects to expectation values of time
ordered products of field operators in the interacting
theory. Therefore, it is not clear how to link the Green functions
to the amplitude of scattering processes.

\section{Summary and discussion}

The algebraic properties of the solution of the free theory of a
noncanonical complex field have been identified. It is a theory of
free particles of two types each one with a different energy-momentum
relation. These relations can be seen as a consequence of nonlinear
deformations of the Poincar\'e algebra in the one particle sectors of
the theory.

The derivation of the path integral formulation of quantum field
theory can be translated to the case of a free noncanonical field
theory and the generalized Lagrangian
version of the path integral formulation can be identified. A
partial mapping to the relativistic theory makes manifest that the
main effect of the noncommutativity is a modification of the time
variable identified as the argument of the field variable in the path
integral formulation. We have identified a Poincar\'e invariance of the
theory in this formulation. Translations in the (physical) time
variable $t$ are symmetry transformations of the free theory which are
not included in the Poincar\'e group of symmetries associated to the
mapping to the relativistic theory. 

The requirement to keep this mapping with the
relativistic theory (and the associated Poincar\'e invariance) to a
theory with interaction leads to
a nontrivial way of introducing interactions in the theory of a
noncanonical complex field. A general expression for the Green
functions in terms of their relativistic  counterparts is given, which
defines a deformation of the interacting relativistic theory of a complex
field parametrized by the noncommutativity length
scale $\theta$.

The non-quadratic terms in the action involve products of fields at
different times (nonlocal interactions). It is not clear whether it is
possible to find an appropriate set of additional variables allowing
to prove the equivalence of the path integral formulation with a
canonical formalism at the level of the interacting theory.
For this reason we can not say at this moment whether
the path integral formulation yields a unitary S-matrix and
the physical interpretation of the theory defined by the
noncanonical deformation of the relativistic Green functions remains
an open problem.

This difficulty may be due to the fact that an interpretation as
a theory of particles is a property of RQFT that may not necessarily
be present in every quantum field theory. In particular, this seems
to be the case of the quantum noncanonical field theory. It is
interesting that different arguments \cite{particle} suggest that
the absence of a direct interpretation in terms of particles might
also be a characteristics of an extension of RQFT trying to
incorporate gravity effects. At this moment however we do not have a
clue of how noncanonical commutation relations could arise as a
trace of the gravitational interaction.

Many of the arguments that we have used in the quantum theory of a
noncanonical complex field can be applied to other attempts to go
beyond relativistic quantum field theory. A discussion along these
lines of the quantum field theory formulation of DSR-like theories and
its relation with canonical
implementations of DSR in position space~\cite{canonicalDSR} will be
presented elsewhere~\cite{CCIM}.

We would like to thank Stefano
Liberati, Florian Girelli, Mikhail Plyushchay and Lorenzo Sindoni for enlighting
discussions. J.I. and D.M. also thank SISSA (Trieste) for
hospitality. This work has been partially supported by CICYT (grant
FPA2006-02315) and DGIID-DGA (grant2008-E24/2). J.I. acknowledges a FPU
grant and D.M. a FPI grant from MEC.

\end{document}